\pdfoutput=1
\documentclass[sigconf,nonacm]{acmart}

\usepackage{amssymb}
\usepackage{dsfont} 

\usepackage{amsmath}
\usepackage{amsfonts}
\usepackage{siunitx} 

\usepackage{booktabs}      
\usepackage{array}         
\usepackage{makecell}      
\usepackage{multirow}      
\usepackage{tabularx}      
\usepackage{threeparttable}
\usepackage{adjustbox}     

\usepackage{xcolor}
\usepackage{tikz}
\usepackage{subcaption}    
\usepackage[most]{tcolorbox} 
\usetikzlibrary{shapes,arrows,positioning,calc,fit,backgrounds}

\usepackage{url}
\usepackage{soul}      
\usepackage{float}     
\usepackage{placeins}  
\usepackage{balance}
\restylefloat{table}

\renewcommand{\arraystretch}{1.1}
\DeclareMathOperator*{\argmax}{arg\,max}

\newcommand{\Dhist}{\mathcal{D}} 

\newcommand{\Ind}[1]{\mathds{1}\{#1\}} 

\newcolumntype{Z}{S[table-format=4.3(4.3)]} 

\newcolumntype{R}{S[table-format=1.4]}

\newcolumntype{C}{S[table-format=1.4(1.4)]}

\newcolumntype{P}{S[table-format=2.1]}

\newcolumntype{L}[1]{>{\raggedright\arraybackslash}p{#1}}

\AtBeginDocument{%
  }

\renewcommand\footnotetextcopyrightpermission[1]{}
\begin{document}
\title{Decision-Theoretic Stopping Rules for Document Screening}

\author{Aaron H.A. Fletcher}
\orcid{0000-0002-4776-066X}
\affiliation{%
  \department{School of Computer Science}
  \institution{University of Sheffield}
  \city{Sheffield}
  \country{United Kingdom}
}
\email{ahafletcher1@sheffield.ac.uk}

\author{Mark Stevenson}
\orcid{0000-0002-9483-6006}
\affiliation{%
  \department{School of Computer Science}
  \institution{University of Sheffield}
  \city{Sheffield}
  \country{United Kingdom}
}
\email{mark.stevenson@sheffield.ac.uk}

\renewcommand{\shortauthors}{Aaron H.A. Fletcher and Mark Stevenson}

\begin{abstract}
Deciding when to stop reviewing the results of a search is a common problem with multiple applications. Existing stopping rules developed within Technology-Assisted Review (TAR) aim to achieve a pre-specified recall target and do not take into account the reason for examining the results, potentially leading to sub-optimal recommendations.  This paper applies decision theory to the problem and uses it to derive three practical stopping policies based on the Expected Value of Perfect Information. The approach is applied to two professional search tasks: patent examining and systematic reviewing. Experiments on CLEF-IP and medical systematic review datasets show that the proposed approach generally produces more appropriate stopping decisions than existing methods, as demonstrated by higher net utility under the evaluated cost and payoff settings.

\end{abstract}



\maketitle

\section{Introduction}

Stopping rules for Technology-Assisted Review (TAR) support reviewers in deciding when they can safely cease examining documents. They have been applied to a range of problems, including e-Discovery in legal IR and systematic reviews of scientific literature, and have been demonstrated to significantly reduce the effort required to screen collections for relevance \citep{cormack_evaluation_2014,cormack_quest_2018,Grossman2016TREC2T,trec_overview_2015}. 
Existing stopping rules aim to ensure that a fixed portion of the relevant documents (known as the {\it target recall}) has been identified. This goal is motivated by the fact that stopping rules were originally developed for scenarios from the legal domain where a litigant may be required to identify relevant documents for disclosure to an opposing party, potentially with statistical guarantees that the target recall has been achieved \cite{cormack2016engineering,lewis_certifying_2021}. Using target recall as the stopping goal has practical advantages: recall is a well-understood metric, and a wide range of standard test collections provides the relevance judgements necessary to measure recall. 
However, the motivation for screening varies according to the task, and this goal is not always appropriate. In some circumstances, it may be apparent that screening can cease before examining enough documents to achieve a given target recall, but in other situations, further screening may still be needed after it has been achieved.

This paper introduces an alternative approach to developing stopping rules by applying decision theory. Decision theory holds that a rational agent takes action only when the expected benefit of that action exceeds its cost and, importantly, provides methods to quantify the expected benefit of acquiring additional information~\citep{von_neumann_theory_2007,savage_1954,briggs_decision_2006}. This approach allows us to develop stopping rules that take into account the motivation for the screening task and the cost of examining additional documents. 
Our objective is to develop rules that guide rational decision making, rather than create models that describe users' behaviour as others have done, e.g. \cite{azzopardi2013query,cooper1973selecting,kraft1979stopping,maxwell2015searching}. We focus on screening ranked lists of documents rather than interactive search, e.g., \citet{azzopardi2011economics}.

 
Our approach assumes that documents are screened to inform a decision, that the decision has consequences, and that examining documents incurs a cost. For example, a patent examiner may want to decide whether to grant or reject an application based on prior art~\citep{roda_clef_2009,mahdabi_queries_2011}. Granting an application for a patent for which there is no prior art or rejecting an application when there is would be regarded as appropriate decisions that lead to positive consequences, but, on the other hand, granting a patent when there is prior art or rejecting one when there is not would be incorrect decisions leading to negative consequences. To help make this decision, the patent examiner has access to the results of a prior art search relevant to the application, but reviewing these results takes time, so there is a cost associated with examining each result. Consequently, the patent examiner seeks to make a correct decision without examining more search results than necessary. The patent examiner only needs to find a single document describing prior art to know that rejecting the application is the correct decision, even if the search results contain many such examples. 

However, the decision to stop may not be straightforward, for example, in systematic reviews when the goal is to compare a new treatment with an existing one to decide whether it should be adopted \citep{higgins_cochrane_2022,institute_of_medicine_us_finding_2011}. If the new treatment is more effective than the existing alternative, adopting it will have positive consequences; if it is less effective, adopting it will have negative consequences, while retaining the existing treatment would be the correct decision. Similar to the patent examiner scenario, the systematic reviewer examines search results to gather evidence on the treatment's effectiveness and aims to minimise the effort required to do so. However, evidence is cumulative in this scenario since each relevant document refines the estimate of the treatment's effectiveness. The systematic reviewer must decide whether to act on the available information, accept its consequences, or examine additional documents at additional cost to obtain more information. 

This paper makes several contributions: formulates TAR stopping as a decision problem; applies decision-theoretic and Value of Information (VOI) principles to derive three practical stopping policies (Greedy, Smooth, Batch); applies the approach to two screening tasks with distinct evidence structures (patent prior-art search and systematic review); and shows that, under the evaluated cost and payoff settings, these policies generally achieve higher net utility than existing recall-centric methods.

\section{Related Work}

The problem of deciding when to stop gathering evidence has been studied extensively across statistics, economics, and information retrieval, including sequential hypothesis testing, optimal search, and stopping models for interactive retrieval~\citep{wald_sequential_1945,stigler_economics_1961,mccall_economics_1970,cooper1973selecting,kraft1979stopping,azzopardi2013query,maxwell2015searching}. Formally, it is a sequential decision problem: an agent observes outcomes, updates beliefs, and decides whether to continue paying for more information or to stop and act. This structure underpins diverse applications, from sequential hypothesis testing in clinical trials~\citep{wald_sequential_1945} to job search in economics~\citep{mccall_economics_1970,stigler_economics_1961}. While the specific state spaces (the information available to the agent) vary, from rank-only information in the secretary problem~\citep{ferguson_who_1989} to unknown reward distributions in multi-armed bandits~\citep{gittins_bandit_1979}, the core challenge remains constant: balancing the cost of exploration against the expected value of exploitation.

Existing approaches generally treat stopping as an optimisation of the search process itself. The goal is typically to maximise the yield of relevant documents relative to the effort required. In some formulations, this optimisation is explicit, using likelihood ratios or bandit indices to predict future yield~\citep{chow_great_1971,gittins_bandit_1979}. In TAR, it is often implicit; methods employ ranking assumptions, effort budgets, or recall targets as proxies for value~\citep{lewis_certifying_2021,moffat_rank-biased_2008}. For instance, the Probability Ranking Principle~\citep{robertson_probability_1977} implies a monotone decrease in the probability of relevance, which knee-point methods exploit to detect when the rate of return diminishes~\citep{satopaa_finding_2011}. However, these methods value information only in terms of ``how many relevant items are found'', ignoring the specific impact of those items on the user's ultimate objective.

A parallel stream of research in health economics values information for its ability to improve consequential decisions. Rather than asking ``Is the next document relevant?'', VOI analysis asks ``Will the next document change our decision?''~\citep{howard_information_1966,claxton_irrelevance_1999}. The Expected Value of Perfect Information (EVPI) quantifies the cost of current uncertainty: how much better, on average, would decisions be if the truth were known? It bounds the maximum benefit any evidence could provide. The Expected Value of Sample Information (EVSI) quantifies the expected gain from a specific, realistic observation—one more document, one more study~\citep{ades_expected_2004,heath_calculating_2018}. Comparing EVSI to screening cost yields a principled stopping rule: continue only when the expected improvement in decision quality exceeds the cost of obtaining it.

\subsection{TAR Stopping Rules}\label{sec:tar_rules}

Current TAR stopping rules focus on the number of relevant documents identified, rather than considering the information each contains or the goal of the screening task. 
Automated stopping strategies based on the observed distribution of relevant documents have been proposed for medical reviews~\cite{passi_2018}, while hypothesis-testing approaches stop when a recall target can be certified at a specified confidence level~\cite{callaghan_statistical_2020}. Improved prevalence estimation has been used to obtain performant target stopping rules~\cite{molinari_sal_2024}, and point process models have been applied to estimate recall from rate functions fitted over the ranking~\cite{stevenson_stopping_2023}. Confidence-based methods like QBCB stop when a recall bound is satisfied, providing statistical certification but ignoring the cost required to achieve it~\citep{lewis_certifying_2021}. Heuristic methods, such as knee-point detection, analyse geometric features in the gain curve that correlate with diminishing returns~\citep{satopaa_finding_2011}. Reinforcement learning approaches have recently been proposed to learn stopping policies from data, but still base their reward functions on recall targets~\citep{bin-hezam_rlstop_2024}. 
These methods have a common objective: achieving a target level of recall with statistical guarantees or minimal overshoot. The present work differs since it optimises the expected utility of the downstream decision that screening informs, yielding stopping rules that adapt to the decision-maker's cost structure and risk profile.

Recent work, expanded here, proposes stopping rules for systematic reviewing that consider the stability of statistical estimates of parameters of interest rather than attempting to achieve a target recall~\cite{Fle26}.


\subsection{Decision-Theoretic Foundations}\label{sec:decision-theory}

Decision theory provides formal tools for reasoning about choice under uncertainty. A decision problem involves an agent selecting an action $a$ from a set $A$ when consequences depend on an unknown state $\omega \in \Omega$. A utility function $u(a, \omega)$ encodes preferences over outcomes, and a belief distribution $p(\omega)$ represents uncertainty about which state of the world is true. Rational choice follows the principle of maximum expected utility: select the optimal action $a^* \in \argmax_{a \in A} \mathbb{E}[u(a, \omega)]$, where the expected utility is defined as $\mathbb{E}[u(a, \omega)] = \sum_{\omega} p(\omega) \, u(a, \omega)$.

The subsequent stopping framework draws on three standard elements from this foundation \cite{von_neumann_theory_2007}.

\paragraph{Utility and commensurability.}
A stopping rule must weigh the benefit of finding relevant documents against the cost of screening. This comparison requires a common scale without which principled stopping is impossible~\citep{briggs_decision_2006}. For example, if screening one document costs £5 in analyst time, and finding a relevant document is worth £200 in avoided litigation, both quantities must be expressed in the same units (here, monetary value) before the stopping criterion can be evaluated. 
Utility provides that scale in our work: a scalar representing how good an outcome is for the decision-maker, expressed in monetary units, time, or domain-specific measures. 

\paragraph{World state and observation history.}
Retrieval stopping involves two distinct types of state. The \emph{world state} $\omega \in \Omega$ is the latent truth the searcher aims to discover—whether invalidating prior art exists, whether a treatment is effective. In many cases, this state cannot be observed directly; it can only be inferred. The \emph{observation history} $\Dhist_d = (y_1, \ldots, y_d)$ is what the agent actually knows at depth $d$: the sequence of relevance labels for examined documents. This history induces a belief distribution $P(\omega \mid \Dhist_d)$, which serves as the basis for both the terminal decision and the choice to continue screening.

\paragraph{Value of information as stopping criterion.}
As discussed in the previous section, EVPI and EVSI provide the machinery for comparing the benefit of continued screening against its cost. The key insight is that stopping becomes rational when even perfect information would not justify the screening expense. 

Many decisions permit evidence collection before commitment. A patent examiner may review additional documents; a systematic reviewer may search for more studies. Each observation can shift beliefs, potentially changing which action is optimal. This introduces another decision layer: should the agent pay to acquire more evidence, or act now?

EVPI quantifies the cost of uncertainty: how much better, on average, would the decision-maker's outcome be if the true world state were revealed before acting? Formally, it is the difference between expected utility under perfect knowledge and expected utility under current beliefs~\citep{howard_information_1966,dwyer_howard_2007}:
\begin{equation}
    \text{EVPI} = \sum_{\omega} p(\omega) \max_{a \in A} u(a, \omega) - \max_{a \in A} \sum_{\omega} p(\omega) \, u(a, \omega).
\end{equation}

No real evidence can exceed this bound. EVSI measures the expected improvement from a specific, imperfect observation—how much better the decision is expected to be after seeing one more document and updating beliefs via Bayes' rule~\citep{ades_expected_2004,heath_calculating_2018}.

These concepts reframe evidence-gathering as investment: acquiring an observation is worthwhile only when its expected value exceeds its cost. When even perfect information would not justify the expense ($\text{EVPI} \leq c$), stopping is certainly optimal. The framework developed in the following section operationalises these principles for retrieval stopping.

\section{Methods}
This section develops a framework for screening tasks where documents are examined to inform a consequential decision. The framework captures the essential structure common to such tasks: uncertainty about whether decision-relevant evidence exists, differing consequences for correct versus incorrect conclusions, and a principled criterion for when to stop screening. This framing is deliberately simplified—the binary world state, while restrictive, captures the structure of many consequential decisions: grant or reject, adopt or withhold, disclose or not. Graded conclusions can be accommodated by introducing a threshold, as is done in this work.

The framework is applied to two contrasting scenarios with different evidence structures: patent prior-art search, where a single document provides decisive evidence, and systematic review screening, where evidence accumulates incrementally.

\subsection{Framework}

Consider an agent screening a ranked list of documents, paying a cost of $c$ per document examined, to inform a binary decision. The agent must ultimately select a terminal action $a \in \{a^+, a^-\}$, where correctness depends on an unknown world state $\omega \in \{\omega^+, \omega^-\}$. At depth $d$, the agent has observed history $\Dhist_d = (y_1, \ldots, y_d)$, where $y_i \in \{0,1\}$ indicates whether document $i$ provides decision-relevant evidence. This history induces a belief
\[
b_d = P(\omega^+ \mid \Dhist_d),
\]
the probability that the positive world state is true after screening $d$ documents. The agent now faces a choice: stop and commit to a terminal action based on current evidence, or continue by paying a cost of $c$ to examine document $d+1$.

It is first necessary to specify what happens when screening ends. The agent, upon stopping, must select one of two terminal actions $a \in \{a^+, a^-\}$, where $a^+$ is appropriate when the true state is $\omega^+$, and $a^-$ is appropriate when the true state is $\omega^-$. The consequences of this choice depend on both the action taken and the true (but unknown) world state, captured by the asymmetric payoff matrix shown in Table~\ref{tab:generic_payoff}.

\begin{table}[h]
\centering
\caption{Generic payoff matrix. The utility $u(a, \omega)$ of an outcome depends on the action taken and the true world state.}
\label{tab:generic_payoff}
\begin{tabular}{lcc}
\toprule
 & $\omega^+$ & $\omega^-$ \\
\midrule
$a^+$ & $+B$ & $-L_r$ \\
$a^-$ & $-L_g$ & $+G$ \\
\bottomrule
\end{tabular}
\end{table}

The payoff matrix defines the utility function $u(a,\omega)$ used throughout the framework. The four parameters have natural interpretations and are derived from domain knowledge: $B > 0$ is the benefit of correctly concluding that decision-relevant evidence exists when it does; $G>0$ is the gain from correctly concluding that no such evidence exists; $L_g>0$ is the magnitude of the loss from wrongly concluding that no evidence exists when decision-relevant evidence was present but not found; and $L_r>0$ is the magnitude of the loss from wrongly concluding that evidence exists based on spurious evidence.
With payoffs identified, the next step is to derive how a rational agent should act upon stopping. As screening progresses, the agent updates this belief using the observations collected so far. The expected utilities of the two terminal actions are:
\begin{align}
\mathbb{E}[u(a^+, \omega) \mid b_d] &= b_d B - (1-b_d)L_r, \\
\mathbb{E}[u(a^-, \omega) \mid b_d] &= -b_d L_g + (1-b_d)G.
\end{align}
 A rational agent selects the action with the highest expected utility. Setting $\mathbb{E}[u(a^+) \mid b_d] > \mathbb{E}[u(a^-) \mid b_d]$ and solving yields a threshold

\begin{equation}
\text{Select } a^+ \text{ if } b_d > q^*, \quad \text{where } q^* = \frac{G + L_r}{B + L_g + G + L_r}.
\label{eq:qstar}
\end{equation}
The agent should select $a^+$ (act on the evidence) whenever the belief exceeds a critical value $q^*$. The threshold $q^*$ encodes the relative utility costs of errors. When $L_g$ is large, wrongly selecting $a^-$ under $\omega^+$ is especially costly. This shifts the threshold downward, so the agent requires less evidence before selecting $a^+$.

Having established what a rational agent does if it stops, the next question it faces is whether it should stop. The answer depends on how much the agent's current uncertainty costs - how much better it could do if it knew $\omega$ with certainty before acting?

With perfect information, the agent would choose $a^+$ when $\omega^+$ (gaining $B$), and $a^-$ when $\omega^-$ (gaining $G$), yielding the expected payoff:
\begin{equation}
\mathrm{EU}_{(PI)}(b_d) = b_d B + (1-b_d)G
\end{equation}

The gap between this clairvoyant payoff and the best achievable under uncertainty is the Expected Value of Perfect Information (EVPI). The form of this gap depends on which action the agent would currently take:
\begin{itemize}
    \item If $b_d>q^*$, agent chooses $a^+$, so $EU^*(b_d) = b_d B-(1-b_d)L_r$
    \item If $b_d < q^*$, agent chooses $a^-$, so $EU^*(b_d) = -b_dL_g + (1 - b_d) G$
\end{itemize}

Here, $EU^*(b_d)$ denotes the expected utility of the optimal terminal action under uncertainty: the best utility an agent can achieve given the current belief $b_d$.

At the threshold $q^*$, these expressions are equal. Below threshold, $b_d(B + L_g) $ is smaller; above, $(1-b_d)(G+L_r)$ is smaller. Combining: 
\begin{equation}
\mathrm{EVPI}(b_d) = \min\bigl(b_d(B + L_g),\; (1-b_d)(G + L_r)\bigr).
\end{equation}

The EVPI quantifies the value of perfect information, whereas examining one additional document yields only imperfect information. Let $\rho_{d+1}$ denote the marginal probability that document $d+1$ provides decision-relevant evidence. Since a single observation cannot be worth more than perfect information, the Expected Value of Sample Information is bounded:
\begin{equation}
\mathrm{EVSI}_d \leq \rho_{d+1} \cdot \mathrm{EVPI}(b_d).
\end{equation}

This bound captures two intuitions: if the next document is unlikely to provide decision-relevant evidence ($\rho_{d+1} \approx 0$), screening has little value; if the decision is already clear ($\mathrm{EVPI}(b_d) \approx 0$), screening is unlikely to be worthwhile regardless of document quality.

The stopping rule follows immediately. The agent continues if and only if the expected information value exceeds the screening cost:
\begin{equation}
\textsc{continue} \iff \rho_{d+1} \cdot \mathrm{EVPI}(b_d) > c.
\end{equation}

What remains scenario-specific is how $b_d$ is updated and how $\rho_{d+1}$ is estimated. The two instantiations differ in the nature of the evidence: existential, where one document can suffice, and cumulative, where evidence aggregates across studies.

\subsection{Instantiations}

Unless otherwise specified, the decision belief is initialised with an uninformative prior $b_0 = 0.5$, corresponding to maximum prior uncertainty between $\omega^+$ and $\omega^-$. The scenarios then differ in how evidence updates this belief and how $\rho_{d+1}$ is estimated.

\paragraph{Patent Prior-Art}
In a patent prior-art search, the decision-maker screens ranked documents to determine whether invalidating prior art exists. In this instantiation, $\omega^+$ denotes that invalidating prior art exists. Thus, the generic belief $b_d$ represents the probability that such prior art exists after screening $d$ documents. The terminal actions are $a^+ = \textsc{Abandon}$, acting as if prior art exists, and $a^- = \textsc{File}$, acting as if no prior art exists. The former is appropriate under $\omega^+$, while the latter is appropriate under $\omega^-$. The payoff parameters represent the consequences of this decision:

\begin{itemize}
    \item B: Benefit of correctly advising against filing when prior art exists (saved filing fees, avoided litigation). 
    \item G: Gain from correctly advising to file when no prior art exists (patent granted, licensing revenue).
    \item \textbf{$L_g$}: Loss from wrongly advising to file when prior art exists (wasted fees, rejection, legal exposure)
    \item \textbf{$L_r$}: Loss from wrongly advising against filing when no prior art exists (missed protection, foregone revenue).
\end{itemize}
Evidence is \textit{decisive}: finding any relevant document proves $\omega^+$ with certainty, so $b_d \rightarrow 1$ immediately upon discovery. When no relevant document has been found, the belief updates via:
\begin{equation}
    b_d = \frac{b_0 \cdot M_d}{b_0 \cdot M_d + (1 - b_0)}, \qquad M_d = \prod_{k=1}^{d}(1 - p^+_k)
    \label{eq:md}
\end{equation}

The term $M_d$ is the probability of observing no relevant documents across all $d$ positions examined, assuming prior art exists. It is the joint probability that document 1 was not relevant, and document 2 was not relevant, and so on through document $d$. Here $p_k^+ := P(y_k = 1 \mid \omega^+)$ is the probability that document $k$ is relevant given that prior art exists, estimated via calibration.

The belief update has an intuitive interpretation: examining high-probability documents and finding nothing is strong evidence against $\omega^+$. If prior art truly existed, it would likely have been found by now. Conversely, missing low-probability documents provides little information. As screening proceeds through the ranking, $M_d$ shrinks and the belief $b_d$ reduces accordingly.

The marginal hit probability $\rho_{d+1} = b_d \cdot p^+_{d+1}$ combines two factors: the probability that prior art exists at all ($b_d$), and the probability the next document is relevant given that it does ($p^+_{d+1}$). Both must be non-negligible for continued screening to be worthwhile.

Two stopping modes emerge: \textit{stop on discovery}, when a relevant document is found ($b_d = 1$, EVPI collapses to zero), or \textit{stop on exhaustion}, when $\rho_{d+1}$ falls below the cost threshold before any discovery.

\paragraph{Systematic Review}

In the systematic review instantiation, the reviewer screens documents to decide whether to adopt or reject a diagnostic test based on whether its sensitivity exceeds a clinical threshold $\theta$. Here, $\omega^+$ denotes that sensitivity exceeds $\theta$. Thus, the generic belief $b_d$ represents the probability that the test is effective under the accumulated evidence. The terminal actions are $a^+ = \textsc{Adopt}$ and $a^- = \textsc{Reject}$. Adoption is appropriate under $\omega^+$, while rejection is appropriate under $\omega^-$. The payoff parameters reflect clinical consequences:

\begin{itemize}
    \item B: Benefit of correctly adopting an effective test (patients receive accurate diagnoses, improved outcomes).
    \item G: Gain from correctly rejecting an ineffective test (resources allocated elsewhere, no false confidence).
    \item \textbf{$L_g$}: Loss from wrongly rejecting an effective test (patients denied beneficial diagnostic).
    \item \textbf{$L_r$}: Loss from wrongly adopting an ineffective test (misdiagnoses, inappropriate treatment).
\end{itemize}

Evidence is \textit{incremental}: each included study contributes data (TP, FP, TN, FN counts) to a meta-analysis, shifting the belief $b_d$ gradually toward or away from the decision threshold. No single document resolves uncertainty. The belief is computed from the posterior distribution over sensitivity produced by the meta-analytic model.

Crucially, only documents with extractable diagnostic data (TP, FP, TN, FN counts) update the belief; other documents, however, topically relevant, do not contribute to the meta-analysis. The marginal hit probability $\rho_{d+1}$ is therefore calibrated against the probability of finding documents that actually shift the estimate.

Unlike patent search, there is no closed-form belief update. The posterior $P(\omega^+|\Dhist_d)$ depends on the specific effect sizes, sample sizes, and quality of the included studies, and is computed via meta-analytic models (e.g., the Reitsma bivariate model for diagnostic accuracy). Each new study with extractable data shifts the posterior by an amount that depends on its content - a large, high-quality study moves beliefs more than a small, noisy one.

The stopping rule from Section~3.1 applies directly: continue if $\rho_{d+1} \cdot \mathrm{EVPI}(b_d) > c$. Screening stops when the continuation region around threshold $q^*$ can no longer be justified: either beliefs have moved sufficiently far from threshold (EVPI shrinks toward zero), or $\rho_{d+1}$ has decayed below the cost-adjusted bound. Unlike patent search, there is no ``stop on discovery'' mode - the decision emerges from accumulated weight rather than a single decisive find.

\subsection{Calibration}

The stopping rule relies on $\rho_{d+1}$, the marginal probability that the next document provides decision-relevant evidence under the scenario-specific definition above. Since raw ranking scores are not probabilities, they must be calibrated. The procedure has two stages. First, within-topic normalisation removes scale differences across queries:
\begin{equation}
    z_d = \frac{v_d - \mu_{\text{topic}}}{\sigma_{\text{topic}}}
\end{equation}
where $v_d$ is the raw score for document $d$, and $\mu_{\text{topic}}$ and $\sigma_{\text{topic}}$ are the mean and standard deviation of scores for that topic. Second, isotonic regression maps normalised scores to probabilities while preserving monotonicity~\citep{zadrozny_transforming_2002,barlow_isotonic_1972}, consistent with the Probability Ranking Principle~\citep{robertson_probability_1977}. The model is fit on pooled training data across topics.

The calibration target differs between scenarios: for patent search, $y=1$ indicates a relevant document; for systematic review, $y=1$ indicates a document with extractable diagnostic data.

\section{Experiments}

The framework is evaluated on patent prior-art search and systematic review screening, which share experimental infrastructure but differ in datasets, utility parameters, and calibration targets. All experiments use standoff stopping rules that operate on fixed, precomputed rankings~\cite{lewis_certifying_2021} to isolate stopping from ranking quality. Methods interleaving selection, training, and stopping (e.g., one-phase active-sampling estimators~\cite{adaptivestopping}) are out of scope, as they couple the stopping decision with the selection loop and induce a different cost model.

\subsection{Datasets}

\subsubsection{Patent Prior-Art}
This scenario uses the CLEF-IP 2009 collection, comprising 1,958,955 patent documents published between 1985 and 2000 \cite{piroi_clef-ip_2021}. Patent documents undergo consolidation to address data inconsistencies across publication stages. For each patent family, a virtual document is constructed by merging title, abstract, description, and claims from all available kind codes (applications and grants). English-language content is selected for processing. To maintain experimental tractability, the search pool is restricted to virtual documents judged relevant to at least one CLEF-IP topic; relevance remains topic-specific within this pool.

The original CLEF-IP topics are drawn from patents rejected between 2001 and 2006 for which invalidating prior art is present in the collection. This chronological separation prevents temporal leakage and mirrors real examination workflows. 
 
These topics all represent patents which should be rejected, so additional valid patents are simulated by corpus sampling. A patent document
$p$ from the collection was selected as a valid topic if: (i)
$p$ does not appear as a relevant document for any topic in the CLEF-IP relevance assessments, (ii) 
$p$ is not itself an existing topic, and (iii)
$p$ contains sufficient textual content for meaningful retrieval. The full text of
$p$ serves as the query, mirroring the query construction for invalid topics. Since these patents were never cited as invalidating prior art for any rejected application in the collection, they lacked prior art within the search pool.
For these topics, all documents are irrelevant by construction. 

The final evaluation set comprises 500 topics representing invalid patent applications from the original CLEF-IP collection and 500 topics representing valid applications from corpus sampling, for both training and test splits (1,000 topics each). This balanced construction is consistent with the uninformative prior used in the instantiation. Probability calibration is performed exclusively on $\omega^+$ training topics, as the calibration target is $P(y_k = 1 \mid \omega^+)$, the probability that document $k$ is relevant given that prior art exists. For $\omega^-$
 topics, $P(y_k = 1 \mid \omega^-) = 0$
by definition. The trained calibrator is then applied to both $\omega^+$
and $\omega^-$ test topics; the belief update mechanism (Equation \ref{eq:md}) correctly handles both world states through the cumulative miss probability $M_d$.

The search pool was ranked using Okapi BM25~\citep{10.1145/2682862.2682863} via \textsc{rank\_bm25} with default parameters ($k_1 = 1.5$, $b = 0.75$, $\epsilon = 0.25$), whitespace tokenisation, and lowercasing.\footnote{
https://github.com/dorianbrown/rank\_bm25} The full-text patent under examination served as the query. Evaluated on $\omega^+$ topics using \textsc{trec\_eval}\footnote{
https://github.com/usnistgov/trec\_eval}, the ranking achieves MAP = 0.353, NDCG = 0.573, and Recall@100 = 0.547.

\subsubsection{Systematic Review}

The corpus comprises 135 diagnostic test accuracy (DTA) outcomes derived from 31 systematic reviews from CLEF eHealth 2017–2019 \cite{kanoulas2017clef,kanoulas2018clef,kanoulas2019clef}. Each DTA systematic review evaluates multiple clinical questions (e.g., test sensitivity under different conditions or patient populations) and each such question constitutes a distinct outcome with its own subset of contributing studies. 

For each outcome, the dataset contains a set of documents retrieved from a Boolean query defined within the systematic review. The CLEF data indicates which of these contain diagnostic accuracy data (i.e. are relevant). For the CLEF 2017 and 2018 datasets this document-level data can be extracted from the LIMSI-Cochrane dataset \cite{norman2018data,norman2019measuring}, but for the CLEF 2019 reviews, which postdate LIMSI-Cochrane, diagnostic counts were extracted directly from Cochrane Library source documents using OCR. Documents are classified as utility-labelled (PMIDs with extractable diagnostic counts for the outcome) or unlabelled (included PMIDs lacking extractable counts). Outcomes were retained only if $\ge5$ utility-labelled studies were available. Evidence is synthesised using the Reitsma bivariate model, producing posterior distributions over sensitivity \cite{reitsma_bivariate_2005}.

The corpus is split into 67 training outcomes (401,625 documents) and 68 test outcomes (312,790 documents). Documents are ranked using Okapi BM25 ($k_1=1.5$, $b=0.75$, $\epsilon = 0.25$), similar to the event-based dataset with systematic review titles as queries, using broad relevance. Evaluation using \textsc{trec\_eval}, Okapi BM25 ranking achieves MAP = 0.099, NDCG = 0.377, and Recall@100 = 0.304 on this dataset.

Calibration quality was evaluated on held-out test sets using the Brier score and Expected Calibration Error (ECE). For the Prior Art Patent scenario, isotonic regression achieved a Brier score of 0.0018 and ECE of 0.0004. For the Systematic Review Scenario, calibration achieved a Brier score of 0.025 and ECE of 0.0002, indicating isotonic regression produces reliable probability estimates for expected utility calculations.

\subsection{Evaluation Metrics}

For each topic $t$, let $d^*_t$ denote the stopping depth, $n_t$ the total number of documents screened, and $\omega_t \in \{\omega^+, \omega^-\}$ the true world state. For patent search, $\omega_t = \omega^+$ if any relevant prior art exists in the collection; for systematic review, $\omega_t = \omega^+$ if pooled sensitivity exceeds threshold $\theta$. Results report mean values across topics. The primary evaluation criterion is net utility, which directly measures the quantity that the proposed framework optimises. Recall, screening cost, decision agreement, and ICER are reported as diagnostic secondary metrics for comparability with prior work and to characterise the trade-offs involved. Six metrics are reported:

\noindent{{\bf Net utility (Net Uti.)}} The primary evaluation metric is per-topic net utility:
\[
U_t = u(a^*_t, \omega_t) - c \times n_t,
\]
where $u(a, \omega)$ is the realised payoff and $a^*_t$ is the terminal action selected by the stopping policy.

\noindent{{\bf Screening cost (Scr. Cost)}} The mean number of documents screened before stopping, $\bar{n} = \frac{1}{|T|}\sum_t n_t$.

For QBCB, which samples documents in addition to traversing the ranked list, sampled documents are included in the cost; documents encountered via both sampling and ranked traversal are counted once.

\noindent{{\bf Recall.}} The fraction of relevant documents retrieved by stopping depth, reported for comparability with prior work. Recall is \textit{not} the optimisation target for the proposed stopping methods.

\noindent{{\bf Decision agreement (Dec. Agr.)}} The proportion of topics where the terminal action matches the optimal action under complete information: $\frac{1}{|T|}\sum_{t=1}^{|T|} \Ind{a^*_t = a^\dagger_t}$, where $a^\dagger_t$ denotes the optimal action given the true world state ($a^\dagger_t = a^+$ if $\omega_t = \omega^+$; $a^-$ otherwise).

\noindent{{\bf Incremental Cost-Effectiveness Ratio (ICER)}}
To quantify the economic trade-off between screening effort and decision quality, the Incremental Cost-Effectiveness Ratio relative to the \textit{Greedy} baseline (the simplest utility-aware stopping rule) is reported \cite{weinstein_critical_1973}. It is calculated as:

\begin{equation}
    \text{ICER} = \frac{\text{Scr. Cost}_{\text{approach}} - \text{Scr. Cost}_{\text{baseline}}}{\text{Dec. Agr.}_{\text{approach}} - \text{Dec. Agr.}_{\text{baseline}}}
\end{equation}

The ICER represents the marginal cost extrapolated to a full unit of accuracy (0 to 1). Practically, an ICER of $V$ implies that gaining 1 percentage point (0.01) of accuracy costs roughly $V/100$ additional documents. Lower positive values indicate efficient scaling, while high values indicate diminishing returns. Negative ICERs indicate dominance, which is attributed to either Dominant (Good), where a lower cost and higher accuracy are achieved, or Dominated (Bad), where a higher cost and lower accuracy are achieved.

\noindent{{\bf Regret (Regret).}} The utility gap between a method and the best-performing evaluated method at each cost level:
\[
\text{Regret}_t = \max_{m \in \mathcal{M}} U_t^{(m)} - U_t
\]
where $\mathcal{M}$ is the set of evaluated stopping policies. This measures how much utility is sacrificed by selecting a given policy from the candidate set. A method with constant competitive regret across cost levels adapts appropriately to varying costs; regret scaling with $c$ indicates failure to adjust screening depth.

\subsection{Proposed Utility-Based Policies}

The decision-theoretic stopping rules from Section~3 yield three operational variants:

\begin{enumerate}
    \item \textbf{Greedy}: Continue if and only if $\rho_{d+1} \cdot \mathrm{EVPI}(b_d) > c$.
    \item \textbf{Smooth ($K$)}: Smooth the probability estimate over the next $K$ documents: $\bar{\rho}_{d+1} := \frac{1}{K} \sum_{j=1}^{K} \rho_{d+j}$. Continue if $\bar{\rho}_{d+1} \cdot \mathrm{EVPI}(b_d) > c$.
    \item \textbf{Batch ($K$)}: Treat the next $K$ documents as a committed batch. Continue if:
\[
\sum_{j=1}^{K} \rho_{d+j} \cdot \mathrm{EVPI}(b_d) > K \cdot c.
\]

\end{enumerate}
Payoffs are normalised such that $B + L_g + G + L_r = 1$. The decision threshold $q^* = (G + L_r)/(B + L_g + G + L_r)$ is varied, so the evaluation and decision-theoretic stopping criteria depend only on $q^*$ and the screening cost $c$, not on the individual payoff values. Three decision thresholds were evaluated, representing an agent's risk profile: \textbf{Aggressive} ($q^* = 0.2$), requiring only weak evidence to act; \textbf{Balanced} ($q^* = 0.5$), with symmetric error costs; and \textbf{Conservative} ($q^* = 0.8$), requiring strong evidence before acting.

Payoffs are set symmetrically: $B = L_g = (1-q^*)/2$ and $G = L_r = q^*/2$, so $q^*$ alone determines both the stopping policy and the realised payoffs, meaning the benefit of a correct decision equals the magnitude of loss from an incorrect one.

\subsection{Baselines}
Four baselines representing diverse approaches to TAR stopping are compared, spanning naive heuristics, statistical certification, geometric detection, and learned policies:

\noindent\textbf{Budget}: Stops after screening a fixed number of documents. This naive baseline represents practitioners who allocate a fixed screening effort regardless of collection characteristics or observed evidence.

\noindent\textbf{QBCB}: A statistical certification method that uses stratified sampling to estimate recall, stopping when a confidence bound exceeds target $\tau$~\citep{lewis_certifying_2021}. It provides probabilistic guarantees but ignores screening cost.

\noindent\textbf{Kneedle}: Detects the geometric knee of the cumulative gain curve, identifying the point of diminishing returns~\citep{satopaa_finding_2011}. The method is cost-agnostic, relying on curve shape rather than utility.

\noindent\textbf{GRLStop}: A reinforcement learning approach that learns stopping policies from training data using recall-based rewards~\citep{bin-hezam_rlstop_2024}. It represents recent data-driven methods but inherits the recall-centric objective from its reward function.

\subsection{Experimental Protocol}

For each stopping approach and cost level $c$, hyperparameters are selected via grid search on the training set to maximise mean net utility. This optimisation is repeated independently for each cost level, so a method may have different optimal hyperparameters at different costs.

\noindent \textbf{QBCB}: $\tau \in \{0.1, \dots, 1.0\}$, $n_{\min} \in \{1, 3, 5, 10, 15\}$, $\alpha \in \{0.1, 0.2, ..., 1.0\}$ was tuned. Results report the mean of 50 replications per topic.

\noindent \textbf{Kneedle}: $\rho_{\min} \in \{ 5, 10\}$, $\beta \in \{10, 100, 500, 1000\}$, $\mathrm{cap} \in \{0.6, 0.75\}$, $f \in \{10, 50, 100\}$.

\noindent \textbf{GRLStop}: Training targets $T = \{0.02, 0.05, 0.1, 0.2, \ldots, 1.0\}$, $W = 100$, $+2$ for achieving the target, $-2$ otherwise, with per-step penalty $-1/W$.

\noindent \textbf{Budget}: screen depth $ \in \{100, 200, 500, 1000\}$. Smaller fixed depths (e.g., $\{1, 5, 10\}$) were excluded because they produce near-zero recall on the systematic review collection, where utility-labelled studies are sparse.

\noindent \textbf{Smooth, Batch}: $K \in \{2, 5, 10, 20, 50, 100\}$.

Policies were then evaluated across the cost values:\\ $c \in \{0.0001, 0.001, 0.002, 0.005, 0.01, 0.02, 0.05, 1.0\}$. For the systematic review scenario, $\theta$ was set to 0.75 (i.e., the decision-maker has sufficient evidence to adopt the diagnostic test if its pooled sensitivity exceeds 75\%).

A natural concern is circularity: policies optimise expected utility, and evaluation reports realised utility. This does not make success trivial. The payoffs $(B, G, L_r, L_g)$, $q^*$, and $c$ are fixed by the task and not tuned to favour any method. Policies act under uncertainty using only $\mathcal{D}_d$ and $\rho_{d+1}$; evaluation uses ground-truth labels. This parallels supervised learning: a loss, training on noisy estimates, evaluation on held-out data.

All experiments use standoff stopping rules operating on fixed, precomputed BM25 rankings. Results, therefore, reflect the interaction between the stopping policy and a specific ranking quality. Stopping quality depends on calibration quality, which in turn depends on the concentration of decision-relevant documents near the top of the ranking. Stronger or different rankers may change both the calibration landscape and the depth at which stopping becomes rational. The present evaluation does not cover active-learning or screening-reranking-rescreening workflows, which couple the stopping decision with the selection loop and introduce a different cost model. Similarly, the cumulative-evidence instantiation (systematic review) relies on a domain-specific meta-analytic model (Reitsma bivariate), which increases deployment cost for new domains. Calibration quality also depends on the availability of labelled training data; low-resource settings may degrade the probability estimates on which the stopping criterion relies. The contribution here is the stopping objective and derived policies for fixed-ranked screening, not a complete treatment of every TAR workflow.

\begin{table}[!t]
\centering
\small
\renewcommand{\arraystretch}{1.1}
\setlength{\tabcolsep}{4pt}
\caption{Patent prior-art performance (c $= 0.001$). Utility-aware policies screen 3–20 documents; recall-centric baselines screen 1,600–2,100, yielding negative net utility despite near-perfect decision agreement.}
\label{tab:patent-omega-icer-final}
\begin{adjustbox}{max width=\linewidth}
\begin{tabular}{l S[table-format=2.4] S[table-format=1.4] r c S[table-format=1.4]}
\toprule
\textbf{Approach} & \multicolumn{1}{c}{\textbf{Net Uti.}} & \multicolumn{1}{c}{\textbf{Recall}} & \multicolumn{1}{c}{\textbf{Scr. Cost}} & \textbf{ICER} & \multicolumn{1}{c}{\textbf{Dec. Agr.}} \\
\midrule

\multicolumn{6}{l}{\textbf{\textit{Panel A: Conservative ($q^*=0.80$)}}} \\
\midrule
Greedy   & 0.2248  & 0.0788 & 3.4    & Ref.    & 0.8910 \\
Smooth & 0.2245  & 0.0784 & 3.3    & 50.0    & 0.8890 \\
Batch    & 0.2243  & 0.1127 & 3.9    & --      & 0.8910 \\
GRLStop  & 0.2081  & 0.2393 & 31.1   & 503.6   & 0.9460 \\
Budget   & 0.1416  & 0.2734 & 100.0  & 1441.8  & 0.9580 \\
Knee     & -1.3722 & 0.2283 & 1621.2 & 15555.8 & 0.9950 \\
QBCB     & -1.8980 & 0.2870 & 2148.0 & 19675.2 & 1.0000 \\
\midrule

\multicolumn{6}{l}{\textbf{\textit{Panel B: Balanced ($q^*=0.50$)}}} \\
\midrule
Smooth & 0.2087  & 0.0673 & 3.3    & \textbf{Dominant}  & 0.9240 \\
Batch    & 0.2059  & 0.2167 & 19.1   & 412.5   & 0.9500 \\
Greedy   & 0.2031  & 0.0851 & 5.9    & Ref.    & 0.9180 \\
GRLStop  & 0.1919  & 0.2393 & 31.1   & 900.0   & 0.9460 \\
Budget   & 0.1290  & 0.2734 & 100.0  & 2352.5  & 0.9580 \\
Knee     & -1.3737 & 0.2283 & 1621.2 & 20977.9 & 0.9950 \\
QBCB     & -1.8980 & 0.2870 & 2148.0 & 26123.2 & 1.0000 \\
\midrule

\multicolumn{6}{l}{\textbf{\textit{Panel C: Aggressive ($q^*=0.20$)}}} \\
\midrule
Greedy   & 0.2005  & 0.0804 & 5.5    & Ref.    & 0.8310 \\
Smooth & 0.2003  & 0.0801 & 5.5    & --      & 0.8300 \\
Batch    & 0.1995  & 0.1143 & 6.1    & \textbf{Dominated} & 0.8290 \\
GRLStop  & 0.1757  & 0.2393 & 31.1   & 222.6   & 0.9460 \\
Budget   & 0.1164  & 0.2734 & 100.0  & 744.1   & 0.9580 \\
Knee     & -1.3752 & 0.2283 & 1621.2 & 9851.8  & 0.9950 \\
QBCB     & -1.8980 & 0.2870 & 2148.0 & 12677.5 & 1.0000 \\

\bottomrule
\end{tabular}
\end{adjustbox}
\end{table}

\section{Results}

\begin{figure*}[bht]
    \centering
    \includegraphics[width=\linewidth]{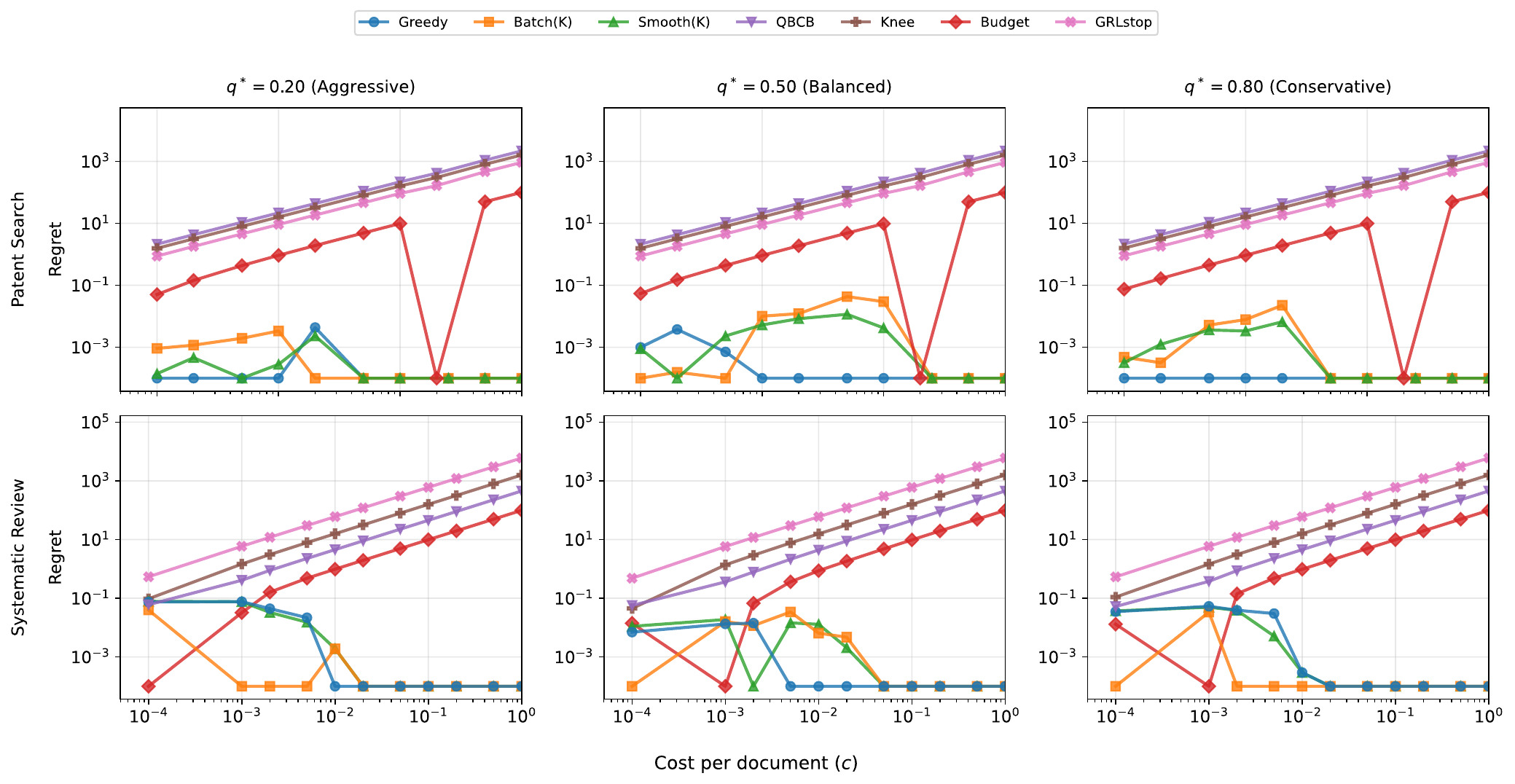}
    \caption{Regret across cost regimes. Regret versus screening cost $c$ on log-log axes. Utility-aware policies (Greedy, Batch, Smooth) maintain near-constant regret across four orders of magnitude of cost; recall-centric baselines (QBCB, Knee, GRLStop) exhibit regret scaling linearly with $c$, reflecting fixed screening depths that ignore cost.}
    \label{fig:regret}
\end{figure*}

Figure~\ref{fig:regret} summarises the central finding: utility-aware stopping policies maintain near-constant regret across four orders of magnitude of screening cost, while recall-centric baselines incur regret proportional to $c$.

\begin{figure*}[htb]
    \centering
    \includegraphics[width=1\linewidth]{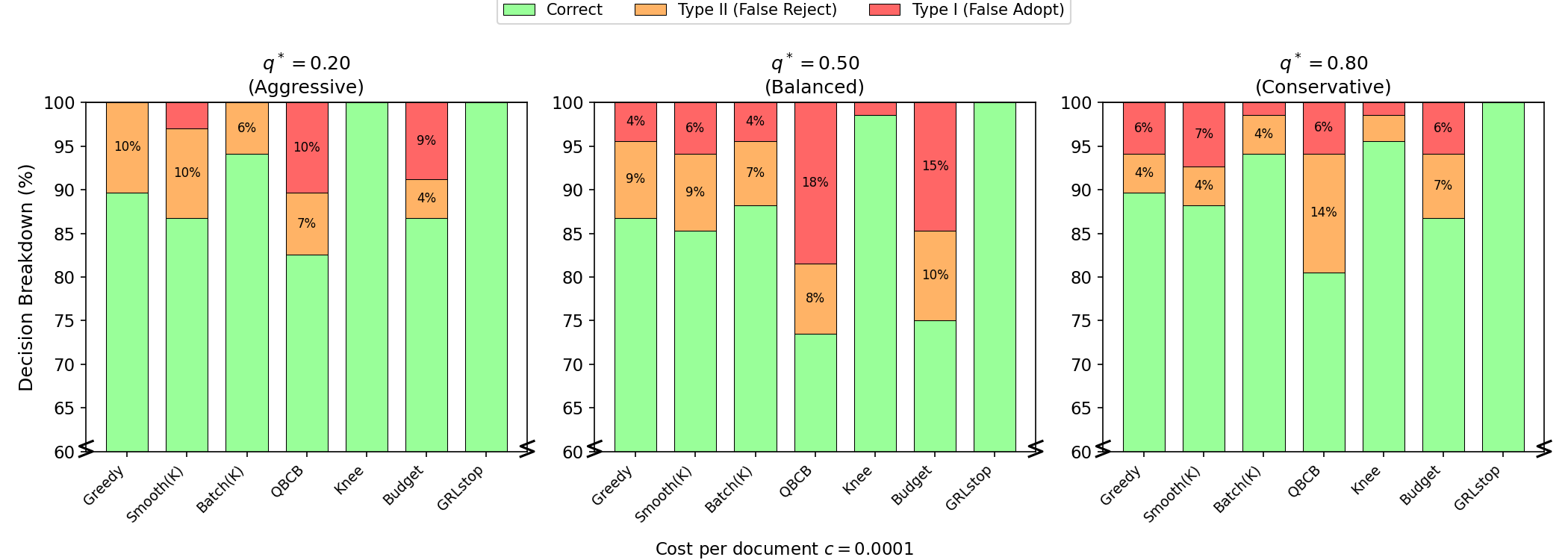}
    \caption{Error type decomposition for systematic review screening ($c = 0.0001$). Type I: false adopt; Type II: false reject. Utility-based methods shift from Type II-dominant errors at aggressive thresholds to Type I-dominant errors at conservative thresholds, consistent with the decision-theoretic trade-off. Baselines show less sensitivity to $q^*$.}
    \label{fig:failure_modes}
\end{figure*}

\subsection{Patent Prior-Art Screening}

Table~\ref{tab:patent-omega-icer-final} presents performance for Patent Prior-Art. The results demonstrate a substantial efficiency gap between the utility-theoretic policy and the recall-centric baselines. Across all risk profiles ($q^*$), the proposed methods (Greedy, Smooth, Batch) achieve positive net utility, whereas the baselines (Kneedle, QBCB) frequently yield negative utility. This difference is driven by screening costs. In the conservative setting ($q^* = 0.8$, Panel A), the Greedy policy achieves a net utility of 0.2248 by screening an average of 3.4 documents. Conversely, QBCB screens an average of 2,148 documents to meet the recall target, yielding a net utility of -1.8980. This massive cost difference outweighs the benefit of the decision itself, rendering the baseline approaches economically irrational in this context. 

The Budget baseline exhibits declining regret at high cost levels in the patent scenario. As $c$ increases, the optimal strategy converges toward minimal screening; Budget's fixed shallow depth (100--200 documents) inadvertently approximates this behaviour, while methods tuned at lower costs may overshoot.

While the Knee/QBCB baselines achieve near-perfect Decision Agreement (0.99--1.00), the ICER analysis reveals the prohibitive price of this perfection. Improving Decision Agreement from 0.89 (Greedy) to 0.99 (Kneedle) in Panel A requires an ICER of over 15,000, implying the agent must screen thousands of additional documents to gain a single percentage point of decision accuracy. In this scenario, the optimal strategy is to stop after identifying fewer than 12\% of relevant documents. This low recall is sufficient to trigger the optimal decision ($a^+$), rendering the remaining 92\% of relevant documents financially redundant.

\subsection{Systematic Review Screening}

Table~\ref{tab:unified-results} details performance for Systematic Review. The utility-based policies again demonstrate superior net utility compared to baselines, though the dynamics differ. The Batch policy generally outperforms the Greedy approach, particularly in the Conservative ($q^*=0.8$) and Balanced ($q^*=0.5$) settings. This suggests that evaluating evidence in batches mitigates the noise inherent in meta-analytic updates, where a single document may not possess sufficient EVPI to justify its cost, but a group of documents might collectively shift the decision.

The GRLStop baseline illustrates the disconnect between retrieval metrics and decision utility. It achieves perfect Decision Agreement (1.0) and Recall (1.0) across all systematic review panels but incurs the highest costs (over 6,000 documents), resulting in the lowest net utility. This confirms that maximising recall is not equivalent to maximising utility and that finding the last few relevant documents may not be cost-effective. 

The Budget baseline achieves the highest net utility in the Aggressive setting ($q^*=0.2$, Panel C), indicating that a simple fixed-depth heuristic may suffice when the decision-maker is willing to act on weak evidence. However, in the Conservative setting (Panel A), Budget falls behind the adaptive Batch policy: Batch's ability to scale screening effort to the evidentiary standard allows it to achieve higher net utility than a fixed stopping depth when stronger evidence is required. These results illustrate that the proposed policies do not dominate the baselines on all metrics or under all conditions; rather, they generally achieve higher net utility under the evaluated cost and payoff settings, while the baselines may achieve higher recall or decision agreement at substantially greater screening cost.

Figure~\ref{fig:failure_modes} decomposes decision errors by type across risk profiles. At aggressive thresholds ($q^* = 0.2$), Type II errors (false reject) dominate for utility-based methods—early stopping occasionally misses sufficient evidence for adoption. As the threshold increases toward conservative ($q^* = 0.8$), the error profile shifts: Type I errors (false adopt) become more prevalent as methods commit to adoption before accumulating contrary evidence. This pattern confirms that the framework trades off error types according to the specified risk profile, rather than minimising total error count. Baseline methods show less systematic variation across $q^*$, consistent with their insensitivity to the decision structure.

A consistent finding across both domains is the inverse relationship between recall and net utility when screening costs are non-trivial. The baselines, which prioritise recall (QBCB) or geometric curve features (Kneedle), implicitly assume that information has infinite value or zero cost. The decision-theoretic framework, by contrast, explicitly models the point at which the marginal value of further screening is outweighed by its cost.

\begin{table}[t]
\centering
\small
\renewcommand{\arraystretch}{1.1}
\setlength{\tabcolsep}{5pt}
\caption{Systematic review screening performance ($c = 0.0001$). Utility-aware policies adapt depth to $q^*$; baselines stop at fixed depths. GRLStop achieves perfect agreement via exhaustive screening at negative net utility.}
\label{tab:unified-results}
\begin{adjustbox}{max width=\linewidth} 
\begin{tabular}{l S[table-format=1.4] S[table-format=1.4] r r S[table-format=1.4]}
\toprule
\textbf{Approach} & \multicolumn{1}{c}{\textbf{Net Uti.}} & \multicolumn{1}{c}{\textbf{Recall}} & \multicolumn{1}{c}{\textbf{Scr. Cost}} & \multicolumn{1}{c}{\textbf{ICER}} & \multicolumn{1}{c}{\textbf{Dec. Agr.}} \\
\midrule
\multicolumn{6}{l}{\textbf{\textit{Panel A: Conservative ($q^*=0.80$)}}} \\
\midrule
Batch    & 0.2226 & 0.3800 & 465.4  & 79     & 0.9412 \\
Budget   & 0.2095 & 0.4775 & 184.1  & 9416   & 0.8676 \\
Greedy   & 0.1876 & 0.3450 & 461.9  & Ref.     & 0.8971 \\
Smooth & 0.1855 & 0.2810 & 365.6  & 6551   & 0.8824 \\
QBCB     & 0.1703 & 0.2540 & 453.0  & 97     & 0.8050 \\
Knee     & 0.1135 & 0.8516 & 1585.9 & 19116  & 0.9559 \\
GRLStop  & -0.3115 & 1.0000 & 6012.4 & 53941 & 1.0000 \\
\midrule
\multicolumn{6}{l}{\textbf{\textit{Panel B: Balanced ($q^*=0.50$)}}} \\
\midrule
Batch    & 0.1290 & 0.3915 & 621.3  & 230    & 0.8824 \\
Greedy   & 0.1220 & 0.3585 & 617.9  & Ref.    & 0.8676 \\
Smooth & 0.1180 & 0.3378 & 584.9  & 2245   & 0.8529 \\
Budget   & 0.1150 & 0.3686 & 100.0  & 4403   & 0.7500 \\
Knee     & 0.0841 & 0.8516 & 1585.9 & 8224   & 0.9853 \\
QBCB     & 0.0721 & 0.2540 & 453.0  & 1241   & 0.7347 \\
GRLStop  & -0.3512 & 1.0000 & 6012.4 & 40745 & 1.0000 \\
\midrule
\multicolumn{6}{l}{\textbf{\textit{Panel C: Aggressive ($q^*=0.20$)}}} \\
\midrule
Budget   & 0.2356 & 0.3686 & 100.0  & 15390  & 0.8676 \\
Batch    & 0.1958 & 0.3419 & 556.6  & 59     & 0.9412 \\
QBCB     & 0.1754 & 0.2540 & 453.0  & 1407   & 0.8253 \\
Greedy   & 0.1608 & 0.3207 & 554.0  & Ref.     & 0.8971 \\
Smooth   & 0.1570 & 0.2998 & 533.3  & 702    & 0.8676 \\
Knee     & 0.1399 & 0.8516 & 1585.9 & 10028  & 1.0000 \\
GRLStop  & -0.3027 & 1.0000 & 6012.4 & 53046 & 1.0000 \\
\bottomrule
\end{tabular}
\end{adjustbox}
\end{table}

\section{Discussion}

The decision-theoretic framework offers two practical advantages over recall-centric baselines. First, it adapts automatically: the stopping threshold is a function of the user's risk profile ($q^*$) and screening cost ($c$), not a fixed hyperparameter. Baselines like Kneedle and QBCB are largely insensitive to cost, failing to curtail screening as $c$ increases. Second, the framework's parameters are semantically meaningful. A practitioner sets $q^*$ by answering ``how confident must I be before acting?'' and $c$ by estimating cost per document—questions with direct operational answers. Baseline methods require tuning abstract parameters (Kneedle's curvature sensitivity $\beta$ and QBCB's confidence level $\alpha$) that lack intuitive interpretations, shifting the burden from domain expertise to hyperparameter search.

The Greedy policy's 7.9\% recall alongside 89.1\% decision agreement (Table~\ref{tab:patent-omega-icer-final}, Panel A) may appear paradoxical when evaluating against recall-centric literature. The gap reflects task structure rather than method failure: for existential search, finding the \textit{first} relevant document suffices to trigger the correct decision; the remaining 92\% of relevant documents are informationally redundant. Low recall is the aim when the goal is decision support rather than exhaustive retrieval.

GRLStop's results with perfect recall and decision agreement, yet yielding the lowest net utility across both scenarios, could be an instance of Goodhart's Law \cite{goodhart_problems_1984}. The reinforcement agent optimised exactly what it was trained to optimise (recall) which does not align with utility. This behaviour is expected as the reward function encoded recall targets which the decision-theoretic framework avoids by construction.

A distinctive property of the EVPI-based criterion is its capacity to justify stopping even when relevant documents likely remain undiscovered. This occurs when the probability-weighted value of that information, discounted by both the likelihood of finding it and its potential to change the decision, falls below the acquisition cost. In such cases, the rational choice is to stop acquiring additional evidence. A patent examiner facing a clearly novel invention need not exhaustively verify the absence of prior art; a systematic reviewer confident in a treatment's ineffectiveness need not screen thousands of additional abstracts.

\section{Conclusion}

This paper applied decision-theoretic principles to formulate TAR stopping as a decision problem and derived three practical stopping policies from an EVPI-based criterion. By formalising the stopping decision as a trade-off between the EVPI and the cost of screening, stopping policies that explicitly optimise for downstream utility, rather than statistical proxies such as recall, were derived. The framework was applied to two scenarios: patent prior-art search (decisive evidence) and systematic review screening (incremental evidence). Experiments on patent prior-art and systematic review datasets showed that the proposed approaches generally achieve higher net utility than existing recall-centric methods. Specifically: (1) Recall-centric baselines incur prohibitive costs to achieve marginal gains in decision accuracy, leading to negative net utility in the evaluated cost regimes. (2) The proposed policies adapt dynamically to varying error costs and screening budgets, maintaining near-constant regret. (3) High recall is neither necessary nor sufficient for high-utility decision making; in patent search, optimal decisions were often reached with less than 12\% recall. 

These findings suggest a shift in TAR evaluation is needed: for consequential screening tasks of the kind studied here, evaluation in terms of downstream decision value may be more informative than recall alone.

\begin{acks}
This work was supported by the UKRI AI Centre for Doctoral Training in Speech and Language Technologies (SLT) and their Applications funded by UK Research and Innovation [grant number EP/S023062/1].
\end{acks}

\bibliographystyle{ACM-Reference-Format}
\balance
\bibliography{sample-base}

\end{document}